\documentclass[sigconf]{acmart}

\AtBeginDocument{%
  \providecommand\BibTeX{{%
    \normalfont B\kern-0.5em{\scshape i\kern-0.25em b}\kern-0.8em\TeX}}}

\copyrightyear{2020}
\acmYear{2020}
\setcopyright{acmlicensed}\acmConference[ETRA '20 Adjunct]{Symposium on Eye Tracking Research and Applications}{June 2--5, 2020}{Stuttgart, Germany}
\acmBooktitle{This is an extended paper for the 2-page abstract accepted at the Symposium on Eye Tracking Research and Applications (ETRA '20 Adjunct), June 2--5, 2020, Stuttgart, Germany}
\acmDOI{10.1145/3379157.3390511}
\acmISBN{978-1-4503-7135-3/20/06}

\setcitestyle{square}

\usepackage{caption}
\usepackage{subcaption}

\begin{document}

\newcommand{\fgsmEpsilon}{\varepsilon} 
\newcommand{\epsStep}{\varepsilon_{s}} 
\newcommand{\epsMax}{\varepsilon_{max}} 
\newcommand{\ltwo}{L_2}
\newcommand{\deltaStep}{\delta_{s}}
\newcommand{\deltaMax}{\delta_{max}}
\newcommand{\alphaStep}{\alpha_{s}} 
\newcommand{\alphaMax}{\alpha_{max}}

\title[Adversarial Attacks on Classifiers for Eye-based User Modelling]{Adversarial Attacks on Classifiers\\for Eye-based User Modelling}

\author{Inken Hagestedt}
\affiliation{\institution{CISPA Helmholtz Center for Information Security,\\ Saarland Informatics Campus}
 }
\email{inken.hagestedt@cispa.saarland}

\author{Michael Backes}
\affiliation{\institution{CISPA Helmholtz Center for Information Security,\\ Saarland Informatics Campus}
 }
\email{backes@cispa.saarland}

\author{Andreas Bulling}
\affiliation{\institution{
University of Stuttgart,\\ Institute for Visualisation and Interactive Systems}
 }
\email{andreas.bulling@vis.uni-stuttgart.de}

\begin{abstract}
An ever-growing body of work has demonstrated the rich information content available in eye movements for user modelling, e.g. for predicting users' activities, cognitive processes, or even personality traits.
We show that state-of-the-art classifiers for eye-based user modelling are highly vulnerable to \textit{adversarial examples}: small artificial perturbations in gaze input that can dramatically change a classifier's predictions.
We generate these adversarial examples using the Fast Gradient Sign Method (FGSM) that linearises 
the gradient to find suitable perturbations.
On the sample task of eye-based document type recognition we study the success of different adversarial attack scenarios: with and without knowledge about classifier gradients (white-box vs. black-box) as well as with and without targeting the attack to a specific class,
In addition, we demonstrate the feasibility of defending against adversarial attacks by adding adversarial examples to a classifier's training data.
\end{abstract}



\begin{teaserfigure}
  \centering
  \includegraphics[trim={0 8.5cm 0 5.4cm},clip,width=\textwidth]{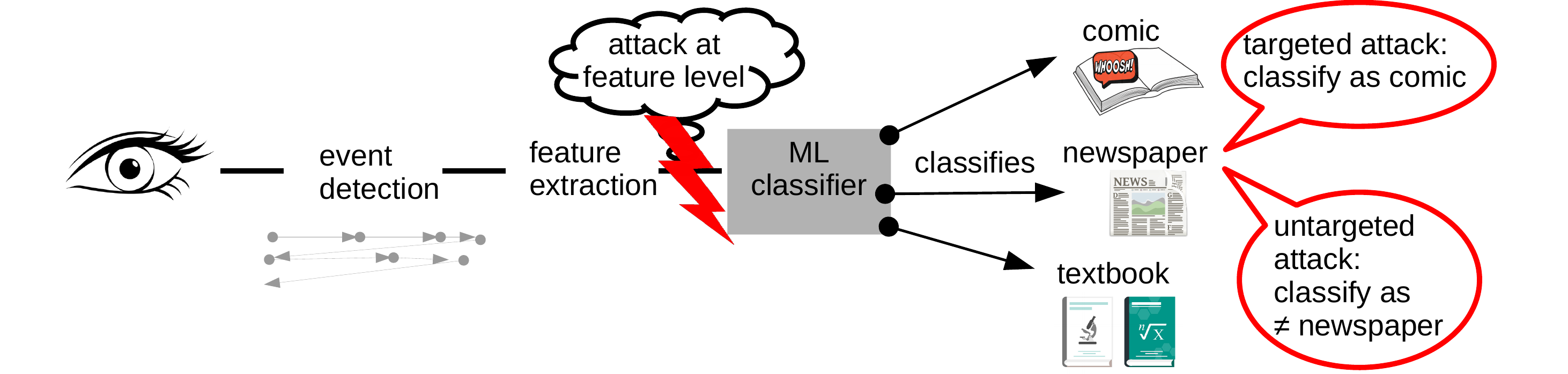}
  \caption{We study adversarial attacks on classifiers for eye-based user modelling at the level of raw gaze data and after feature extraction, with and without targeting a specific class, as well as with and without knowledge about the internal classifier gradients. We do so on the sample task of document type classification from eye movements during reading.}
  \label{fig:teaser}
\end{teaserfigure}

\maketitle

\section{Introduction}
In recent years, advances in mobile eye tracking \cite{tonsen17_imwut,kassner14_ubicomp} as well as learning-based gaze estimation and eye contact detection using off-the-shelf cameras \cite{zhang19_pami,zhang17_uist} have spurred research on eye-based user modelling.
That is, the use of machine learning to predict various (potentially previously hidden) user characteristics from eye movements, such as users' activities \cite{bulling2010eye}, cognitive processes and states \cite{sattar15_cvpr,bulling2014cognition}, or even personality traits~\cite{hoppe2018eye}.
Combined with the continuing integration of eye tracking into head-mounted virtual and augmented reality headsets, eye-based user modelling promises a range of exciting new applications that may already become possible in the near future, such as mental health monitoring \cite{vidal12_comcom} or life logging and quantified self \cite{kunze2015quantifying,kunze2013activity}.

However, with ever-more widespread integration and thus availability of gaze data comes an ever-increasing risk of misuse and attacks on users' privacy.
Examples for potential adversaries are headset manufacturers trying to obtain
personal information about consumer interests or preferences, malicious applications running on the host computer to which the headset is connected that spy on users' activities, or third parties trying to exploit eye gaze data that is available publicly or that has been obtained in a targeted attack.
Despite all of these diverse threats, research has so far mainly focused on ocular biometrics \cite{nigam2015ocular} or secure user authentication using eye movements \cite{holland2011biometric}.
Researchers have only recently started to study some of these threats and proposed first solutions to making eye tracking privacy-aware -- both at the hardware \cite{steil2019privaceye} and software \cite{steil2019privacy,john2019eyeveil,liu2019differential} level.

This work contributes another building block to this emerging research field of \textit{privacy-aware eye tracking} by studying, for the first time, adversarial attacks.
That is, the process of creating small artificial perturbations in gaze data input, better known as \textit{adversarial examples}, that, if fed to a classifier, can dramatically change the classifier's predictions.
Adversarial examples and attacks have only fairly recently started to being studied in the intersecting research field of computer vision and security 
\cite{papernot2018sok} but have not yet received any attention in the eye tracking community.
We study adversarial attacks on classifiers for eye-based user modelling for the sample task of document type recognition from eye movements during reading \cite{kunze2013know}.
We picked this task given that reading is a truly pervasive activity, widely studied in different fields including eye tracking research, and has been the subject of a recent study on using differential privacy for privacy-aware eye tracking \cite{steil2019privacy}.

We study the feasibility and performance of this attack in different scenarios that we carefully chose to represent real-world use cases (see \autoref{fig:teaser}): Attacks with and without knowledge about the internal classifier gradients (white-box vs. black-box) as well as with and without targeting the attack to a specific class.
To create adversarial examples that are suitable to attack classifiers for eye-based user modelling in the white-box model, we use a recent gradient-based method -- the Fast Gradient Sign Method (FGSM)~\cite{goodfellow2014explaining}.
In a nutshell, FGSM computes the classifier's gradient and iteratively perturbs a particular sample until it crosses the class decision boundary and is misclassified (see \autoref{fig:fgsmExplanation}).
For the black-box attack model, where no gradients are available, 
we propose a simple method that randomly perturbs single data points in the stream of raw gaze data such that the sample is misclassified.

The specific contributions of our work are three-fold.
1) We are the first to study the vulnerability of state-of-the-art classifiers for eye-based user modelling to adversarial attacks.
2) We conduct comprehensive evaluations that provide detailed insights into different adversarial attack scenarios representing real-world use cases.
Furthermore, we explore the feasibility of defending against adversarial attacks by using the same gradient-based method to generate adversarial training data.
3) We discuss the obtained findings and derive a set of recommendations for researchers and practitioners working on eye-based user modelling. 

 
\section{Related Work} 

Our research is related to previous works on 1) predicting user characteristics from eye movements, 2) privacy-aware eye tracking, and 3) the creation and use of adversarial examples.

\subsection{Predicting User Characteristics from Eye Movements}
Eye-based user modelling, i.e. the computational task of predicting diverse user characteristics and general user context from eye movements \cite{bulling11_pcm}, has gained significant attention in recent years.
The field was pioneered by Bulling et al.~\cite{bulling2010eye,bulling12_tap} who demonstrated that different office activities, including reading, could be robustly detected from eye movements alone in both stationary and mobile settings.
They also introduced a large set of new eye movements features that cover the main eye movement types (saccades, fixations, and blinks) and that we use in this work for document type classification from eye movements during reading.
As such also closely related is the work by Kunze et al.~\cite{kunze2013know} who showed that the type of document being read could be predicted from eye movement data.
In later work, Bulling et al.~\cite{bulling2013eyecontext} and Steil et al.~\cite{steil2015discovery} demonstrated that high-level contextual cues -- e.g. social interactions, focused work or whether a person was inside or outside -- could be extracted from day-long recordings of eye movements in both a supervised and unsupervised fashion.

In addition to different activities, the eyes are also a rich source of information on human cognition and user characteristics difficult or impossible to assess using other modalities \cite{bulling14_pcm}.
Early work by Hess et al.~\cite{hess1960pupil} showed that a person's interests are reflected in eye movements, while Matthews et al. demonstrated that the eyes also reveal users' cognitive load~\cite{matthews1991pupillary}.
Eye movements can also differ for women and men, e.g. when looking at faces~\cite{sammaknejad2017gender}.
More recent work has demonstrated that visual search intents, i.e. targets of visual search, can be robustly inferred \cite{sattar15_cvpr,sattar17_iccvw,zelinsky2013eye, jang2014identification} and even visually decoded similar to a photo-fix \cite{sattar17_arxiv} by combining information on users' eye movements and the underlying image.
Recent work has even shown that eye tracking data collected during everyday activities allows to infer personality traits~\cite{hoppe2018eye}. 
Finally, eye movements are linked to a range of potentially highly sensitive mental health issues \cite{vidal12_comcom}, such as Parkinson's~\cite{kuechenmeister1977eye} or schizophrenia~\cite{holzman1974eye}.

All of these works demonstrate that rich but also highly sensitive personal information about who we are, what we do, and how we think can be inferred from eye movements. 
At the same time, people typically do not consciously control their eyes and are, as such, unaware of what they reveal~\cite{steil2019privacy}.
Taken together, these observations point at the significant risks and therefore also at the urgent need to not only better understand possible attack vectors on methods for eye-based user modelling, including the classifiers themselves, but also to develop methods to protect from these attacks.
Adversarial attacks, one of these key attack vectors, has never been studied in the context of eye tracking before.

\subsection{Privacy-aware Eye Tracking}

Privacy and security has only recently started to being investigated in eye tracking research.
Liebling et al.~\cite{liebling2014privacy} were the first to summarize which private information can be extracted from the eyes and discussed several approaches for how to protect privacy. 
One of these ideas was implemented by Steil et al.~\cite{steil2019privaceye} who developed a physical shutter for the scene camera of mobile eye trackers. The system detected privacy-sensitive situations, automatically closed the shutter and re-opened it when eye movement analysis showed that the user moved out of the privacy-sensitive situation or activity. 
Another approach to preserve privacy is the application of noise.
While Steil et al.~\cite{steil2019privacy} applied noise at the feature level using a differential privacy method and showed that gender and identity could be protected while utility for document type classification could be preserved, Liu et al.~\cite{liu2019differential} noised aggregated gaze heat maps.
Our work adds a novel attack vector to the emerging field of privacy-aware eye tracking by, for the first time, attacking the classifier instead of the data.

\subsection{Adversarial Examples}

Research on adversarial examples started with pioneering work by Lowd and Meek~\cite{lowd2005adversarial} and Dalvi et al.~\cite{dalvi2004adversarial}. 
With the rise of neural networks, the issue of adversarial examples has started to receive a lot of additional attention. 
We refer the reader to Papernot et al.~\cite{papernot2018sok} for an extensive overview of the current state of the art. 

We focus in our work on test time attacks, i.e., the attacker does not manipulate the training process but aims to change classifications at test time. 
In this setting, researchers generally distinguish between white-box and black-box attacks.
For white-box attacks, the trained classifier and all its components are available to the attacker, most importantly allowing her to compute the classifier's gradients.
These are used in gradient-based attacks, such as the Fast Gradient Sign Method (FGSM) \cite{goodfellow2014explaining} that we also use in this work. 
The currently best performing attack was developed by Carlini and Wagner~\cite{carlini2017towards}, 
however, it is only applicable to neural networks. 

On the other hand, black-box attacks assume the attacker can only query the classifier, but not inspect its components such as support vectors or weights of a neural network. 
Attacks in this black-box model differ in what they assume about the output format: 
While the HopSkipJump attack by Chen et al.~\cite{chen2019hopskipjumpattack} only requires the class label and are therefore called decision-based, 
other attacks such as the Zeroth-order optimization attack~\cite{chen2017zoo} need the scores for each individual class and are referred to as score-based attacks. 
Additionally, Papernot et al.~\cite{papernot2016transferability} studied the transferability of adversarial examples  and found that substitute models generated by the adversary can be attacked in the white-box-model and 
the resulting examples are often adversarial for the target classifier as well. 

According to Papernot et al.~\cite{papernot2018sok}, the problem of defending a classifier against adversarial attacks is not yet solved. 
We study one proposed method, adversarial training, where we re-train the target classifier on adversarial examples added to the training data. 
This method was introduced by Szegedy et al.~\cite{szegedy2013intriguing} and we opted for this due to its simplicity. 

In summary, while adversarial examples have been explored in the machine learning and security research communities, we are the first to study adversarial attacks and defense in the context of eye tracking and more specifically for eye-based user modelling.


\section{Creating Adversarial Examples for Gaze Data}

We model an attacker who knows the preprocessing pipeline and attacks at feature level. 
The targeted classifier might be known (white-box) or a suitable surrogate classifier can be trained.

To create adversarial examples, we used the Fast Gradient Sign Method~\cite{goodfellow2014explaining} (FGSM), a state-of-the-art method implemented in the Adversarial Robustness Toolbox~\cite{art2018}.
As opposed to other state-of-the-art attacks like Carlini and Wagner~\cite{carlini2017towards} or JASMA~\cite{papernot2016limitations}, FGSM does not rely on neural network structures but only on gradients that are well-defined for SVM with RBF kernel. 
Notice that RF did not have gradients and can therefore not be attacked by FGSM directly, only indirectly by transferability. 

FGSM computed the gradient and perturbed the sample to move it along this direction. 
FGSM can be performed in two modes: standard and minimal. 
While the former implements FGSM as described in the paper by Goodfellow et al.~\cite{goodfellow2014explaining}, 
the minimal mode repeatedly computes a growing adversarial perturbation until the sample is misclassified or a maximal perturbation is reached. 
Because the minimal attack leads to smaller perturbations in general, we applied this form of attack. 

\begin{figure}[t]
\includegraphics[trim={6cm 4.5cm 0 5.5cm},clip,width=0.45\textwidth]{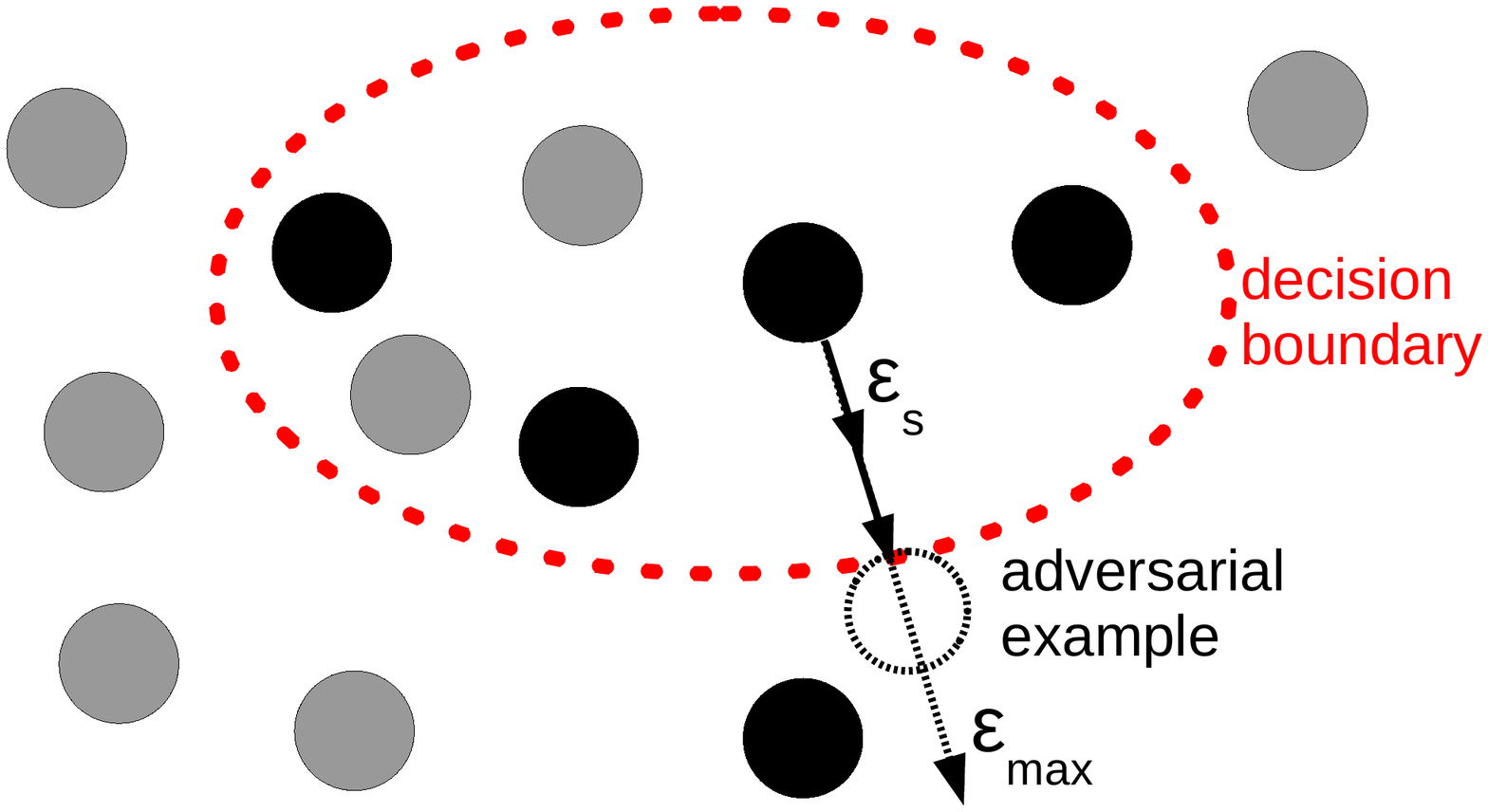}
\caption{Illustration of how FGSM generates adversarial examples in 2D for a simple 2-class classification problem (black vs. gray points). 
The goal of the attack is to cross the decision boundary (red dotted line).
FGSM computes the gradient and moves the point into this direction as shown with the black arrows. 
We used the minimal mode such that the data point is perturbed by multiples of $\epsStep$ iteratively until it is misclassified (here shown as dashed circle) or a maximal perturbation of $\epsMax$ is reached. 
In the example shown here, perturbation by twice $\epsStep$ is sufficient and it is not necessary to perturb the point by the maximal allowed amount $\epsMax$.}\label{fig:fgsmExplanation}
\end{figure}

The minimal FGSM attack has three hyperparameters that need to be optimised: the norm to measure perturbations, the perturbation per step ($\epsStep$), and the maximal perturbation ($\epsMax$). 
\autoref{fig:fgsmExplanation} visualizes FGSM with these hyperparameters. 
We used the $\ltwo$ norm to measure perturbations, which is often used for the generation of adversarial examples and can remain small if there are many small changes to the features~\cite{carlini2017towards}. 

We evaluated two different methods to find the hyperparameters for FGSM, concretely, the maximal perturbation $\epsMax$ we allowed for adversarial examples. 
The general $\epsMax$ was chosen such that the average accuracy over multiple participants was lowest, this corresponds to an attack without prior knowledge about the target's data.
We chose the person-specific $\epsMax$ by determining the maximal perturbation that was needed for the lowest accuracy for each person individually. 
Given that the lowest accuracy might not be necessary in all use-cases, we additionally evaluated how much perturbation was needed to drop the accuracy to chance level.

In order to evaluate the untargeted attack, we perturbed all samples independently of their ground truth label and the goal was to misclassify the sample into any of the other classes. 
In contrast, targeted attacks were evaluated on samples of one class only and the goal was to perturb them such that they were classified as one specific other class. 
Additionally, we studied transferability by evaluating the adversarial examples 
generated for SVM on a random forest classifier fitted on the same data.
The SVM serves as surrogate classifier for the random forest.


\section{Evaluation}

\subsection{Dataset}\label{data-preprocessing}
In this work we used the public dataset by Steil et al. that included recordings of 20 participants reading three different types of document (comic, newspaper, textbook) in virtual reality~\cite{steil2019privacy}. 
Eye movements were recorded using the eye tracking integration from 
Pupil Labs at 30Hz~\cite{kassner14_ubicomp}.
The data includes detected eye positions in x- and y-direction in the range of [0,1], the estimated pupil diameter, the timestamp and how confidently these estimates were detected. 

We first detected fixations using a dispersion-based algorithm.
A fixation was detected if x- and y-positions of gaze samples were within a radius of 0.05 for at last three consecutive frames, i.e., 0.1 seconds and x- and y- positions were not both equal to zero. 
If x- and y-positions were both equal to zero and the confidence was zero for at last three consecutive frames, we recognized a blink. 
Notice that if the eye tracker could not detect the eyes, it returned zero for both x- and y-position as well as confidence. 
We assumed saccades between two fixations, or between fixations and blinks.
We excluded saccades between fixations and blinks if there is not at least one frame in between to avoid situations where a fixation was directly followed by a blink and a saccade between fixation and blink is recognized.
We followed the same approach for blinks directly followed by a fixation.  

In a second step, we extracted a total of 52 high-level features from these basic eye movements as described by Bulling et al.~\cite{bulling2010eye}. 
These features included the ratio of the detected features, means and variances of durations and pupil diameters as well as wordbook features that encode sequences of up to four saccades.
\autoref{tab:features} explains these features in detail.
We extracted these features using a sliding window with the window size being one of the to-be-optimised hyperparameters of our method.
Additionally, we added two more features from Kunze et al.~\cite{kunze2013know}, namely,
the euclidean distance between the 5\% and 95\% quantile of fixation coordinates and the slope of saccade directions using linear regression over fixations. 
These features give a general estimate of the reading direction and distance covered during the time window. 

\subsection{Classifier Training}

\paragraph{SVM Training}

We studied the computational task of recognising different document types from eye movements during reading. 
Support vector machines (SVM) 
are a widely used classifier in the literature on eye-based user modelling~\cite{bulling2013eyecontext,bulling2010eye,karessli2017gaze,sattar15_cvpr}.
We trained with leave-one-person-out cross-validation
using sklearn's SVM implementation~\cite{sklearn_api}.
The optimal window size was selected based on validation accuracy of 200 samples per participant and document type. 
We found a window size of 45 seconds to perform best. 
Notice that we did not tune the penalty parameter $C$ on the error term but kept its default parameter 1.0.
We also did not tune the RBF-kernel hyperparameter $\gamma$ that controls the locality of the kernel and used its default value of $\frac{1}{54}$. 
We report the accuracy on the remaining test samples (about 600 samples) per participant and document type for the selected window.

\paragraph{Random Forest Training}

Decision-tree based classifiers such as random forest (RF) are also used in the eye tracking literature,  e.g., in Kunze et al.~\cite{kunze2013know} even though less frequently than SVMs. 
We study random forests as an example of adversarial example transferability across different types of classifiers. 
Similarly to before, we selected the best window size using leave-one-person-out cross-validation on the same data split into training, validation, and test data. 
Again, we found a window size of 45 seconds to perform best.
We evaluated for the key hyperparameters, namely the number of trees (100, 50, 10, 200) and the number of samples per leaf (50, 10, 100, 5).
Both measures are important to avoid overfitting. 

\subsection{Evaluation Metrics}
We evaluate our attacks in terms of success, i.e., 
document type classification accuracy on adversarial examples,
and in terms of changes induced. 
In order to evaluate the effect of the resulting perturbations, 
we compared the average distance between benign test samples to the distance between the benign and resulting adversarial point at feature level.
We used again the $\ltwo$ distance metric as for the FGSM optimization. 
In case the average distances between benign and adversarial points were higher than the average distances between benign samples, we cannot call the adversarial perturbations ``small'' and conclude an attack significantly deteriorates data quality. 
To show the distances to adversarial examples are significantly lower than between benign samples, 
we applied scipy's implementation~\cite{jones2001scipy} of Welch's t-test, a test whether the two distributions have the same average value, but that does not assume identical variances. 

\section{Results}

\begin{figure}[t]
\includegraphics[width=0.45\textwidth]{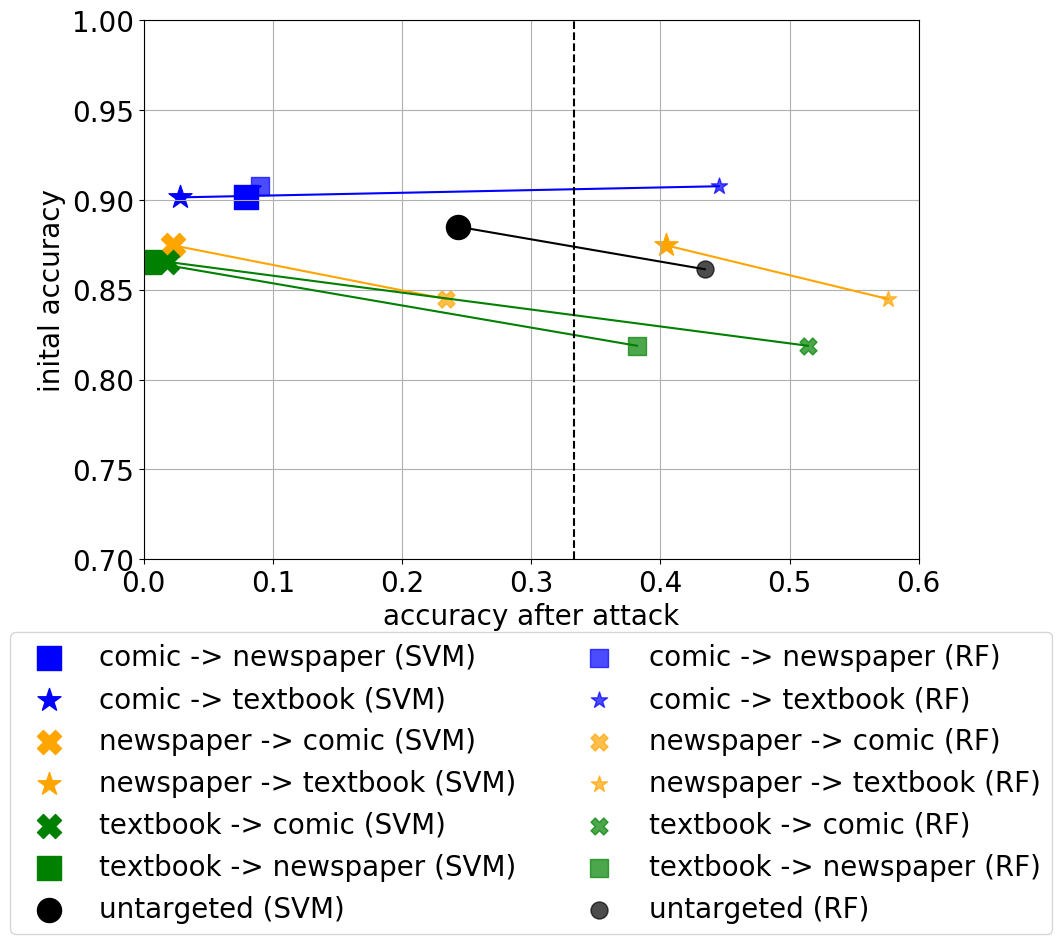}
\caption{Accuracies after attacks on SVM with FGSM (larger markers) and transfer to RF (smaller markers).
The y-axis displays the initial accuracy on benign data and the x-axis the performance on adversarial examples. 
Different colors and markers show different document types under attack.
The dashed black line visualizes the chance level.}\label{fig:fgsmAccuracy}
\end{figure}

\begin{figure}[t]
    \begin{subfigure}{\columnwidth}
    \includegraphics[width=0.9\columnwidth]{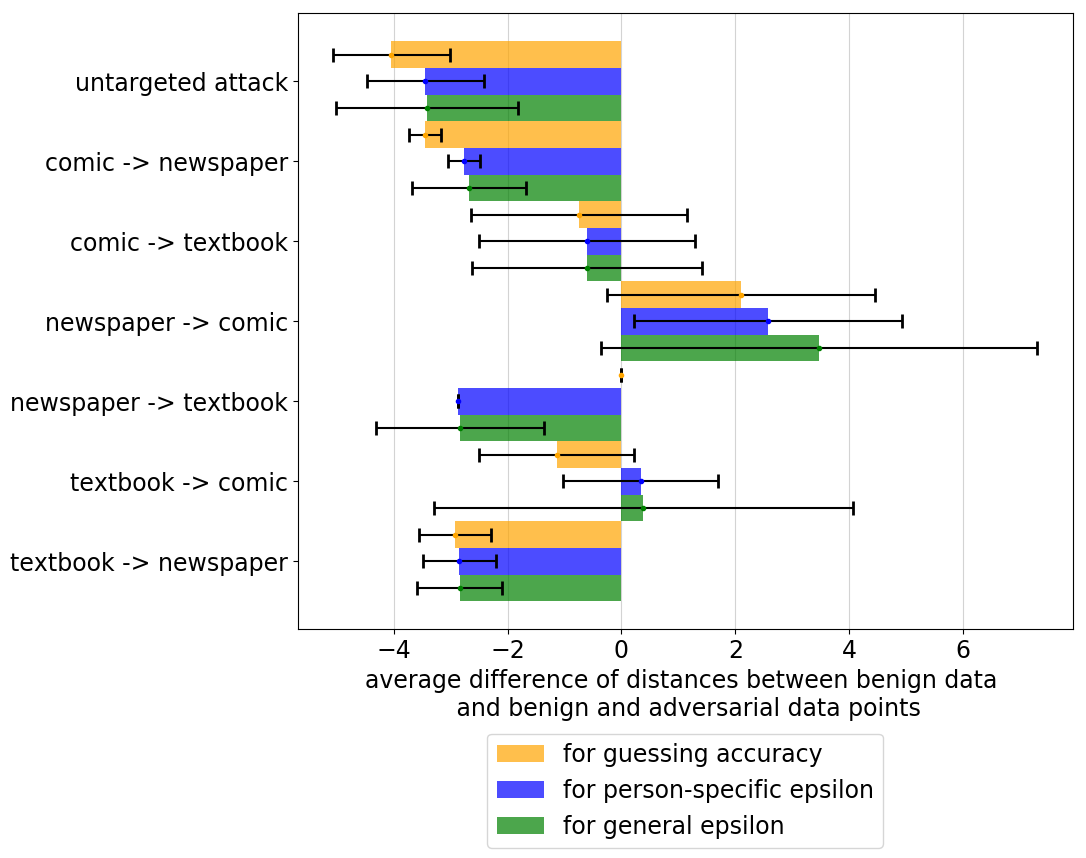}
    \caption{Difference of distances after untargeted and targeted attacks on SVM with FGSM.}\label{fig:fgsmSVMdistances}
    \end{subfigure}

    \begin{subfigure}{\columnwidth}
    \includegraphics[width=0.9\columnwidth]{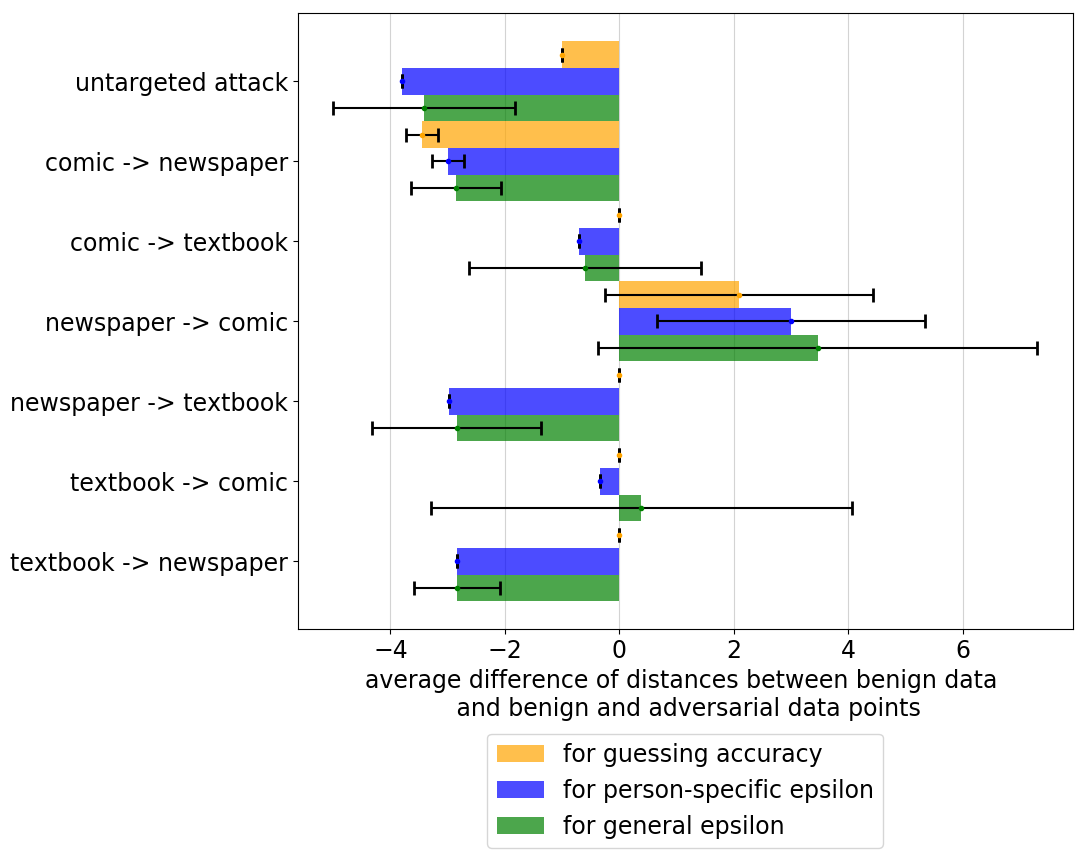}
    \caption{Difference of distances after untargeted and targeted attack on SVM and transfer to RF.}
    \label{fig:fgsmRFdistances}
    \end{subfigure}
\caption{The x-axis shows the average distance of adversarial examples to their original benign point minus the average distance between benign points, bars towards the left indicate the adversarial example is closer to the benign point than the distances naturally occurring between test points, bars towards the right indicate the perturbations exceed these test point distances.
Error bars visualize the standard deviation computed over the 20 participants.
The y-axis shows untargeted attacks and the different targeted attacks that perturb samples from the first given class to be mistaken as the second class.}
\end{figure}

\noindent{\textbf{Attack Success:}}
\autoref{fig:fgsmAccuracy} shows the results of the targeted and untargeted white-box attacks on SVMs using FGSM. 
The accuracy after attack is 
well below chance level of $\frac{1}{3}$
with one exception, the targeted attack to misclassify newspaper as textbook where the accuracy is only around 0.4. 
But even in this case, the accuracy decreased massively from initially between 0.8 and 0.9. 
We conclude that the SVM is vulnerable to our attack. 

We observe in \autoref{tab:results} that the accuracy of general and person-specific choice of $\epsMax$
are very close and therefore plotted the general choice of $\epsMax$ in \autoref{fig:fgsmAccuracy}.
That means, it is not necessary to know data of the target before mounting a successful attack. 

Additionally, we compare the different target classes in~\autoref{fig:fgsmAccuracy}. 
It is easiest to perturb textbook samples such that they look like newspaper readings to the SVM, but textbook examples are also relatively easy to turn into the other class, comics. 
For comic, it is easier to trick the SVM into classifying it as textbook as as newspaper. 
The newspaper examples are hard to turn into textbook samples, but can be misclassified as comic much easier. 
This is surprising because the reverse direction, perturbing a textbook example such that it is misclassified as newspaper, is the easiest in our study. 

\noindent{\textbf{Distance Evaluation:}}
Next, we evaluate the consequences of our adversarial perturbation in terms of euclidean distance. 
We first measured the euclidean distance between all test samples before the attack.
This naturally occurring distance between samples was compared to the distances between test samples and their corresponding perturbed version. 
\autoref{tab:results} reports the resulting distances, and we plotted the difference of distances in \autoref{fig:fgsmSVMdistances}, i.e., average distance after attack minus the average distance of benign samples. 
A value positive value shows that the distance between benign and respective perturbed point is larger than the distance between benign points on average. 
The different coloured bars in \autoref{fig:fgsmSVMdistances} show different ways to select the FGSM hyperparameter $\epsMax$. 
Additionally, we selected the smallest $\epsMax$ such that on average over all participants, an accuracy of 0.3 is reached. 
Notice that this was not always possible and we show no orange bar in that case.
If the goal is only guessing accuracy, smaller perturbations often suffice.
We observe that in most cases, on average the distance between original and perturbed point is smaller than the average distance between benign test points. 
The p-values of Welch's t-test between the two distributions of distances are below 0.01 except for the targeted attacks misclassifying comic as textbook and textbook as comic, respectively.
This demonstrates that the perturbations we computed are indeed mostly ``small''.

\noindent{\textbf{Transferability:}}
Finally, we study whether the perturbations we computed carry over to a different family of classifiers, namely, RF classifiers. 
For that, we used the adversarial examples against the SVM computed as before but classified them with the RF, ~\autoref{fig:fgsmAccuracy} visualizes the results. 
Initially, SVM and RF had similar accuracy without the presence of adversarial examples.
We observe that the accuracy drops after the attack, but only rarely below guessing accuracy for RF. 
That means, the decision boundaries between SVMs and RF are similar enough for some samples to be transferable, however, not so similar that all of them carry over. 
We show the distances' evaluation again in~\autoref{fig:fgsmRFdistances} for RF. 
The distances are quite similar to those observed for the SVM, but in some cases higher. 
In these cases, a larger perturbation that was not necessary for SVM led to misclassification in the RF. 
We conclude that knowledge on the type of classifier does increase attack accuracy, however, hiding the type of classifier does not mitigate all attacks. 

\subsection{Defense with Adversarial Training}\label{sec:defense}
\begin{figure}[t]
\includegraphics[width=\columnwidth]{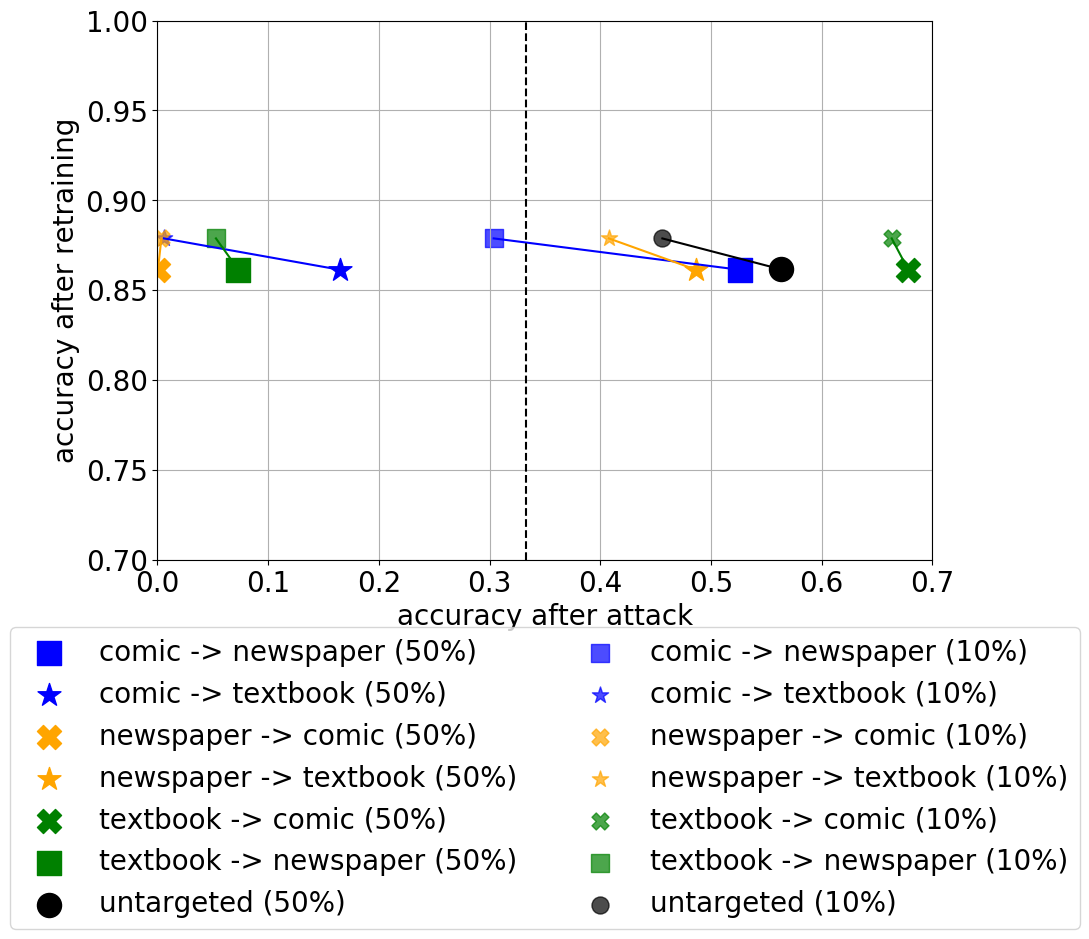}
\caption{The y-axis shows the accuracy after retraining without adversarial examples and the x-axis shows the performance on adversarial examples crafted for the retrained SVM. 
Different colors and markers show different document types under attack.
Retraining with 50\% of the training data is displayed with larger markers, retraining with 10\% with smaller markers.
The dashed black line visualizes the chance level.}\label{fig:retrainAcc}
\end{figure}
\begin{figure}[t]
\includegraphics[width=0.45\textwidth]{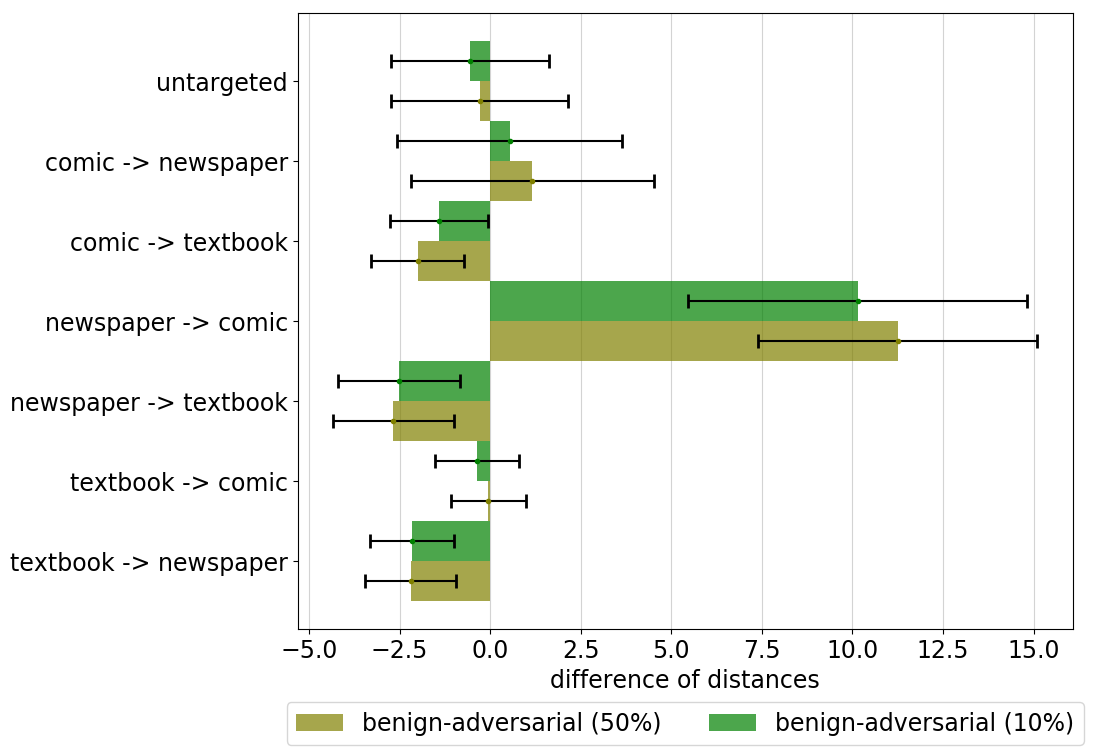}
\caption{Difference of distances after untargeted and targeted attack on SVM that was retrained with adversarial examples.
The bars show the difference between the average distances between benign test points and between benign and respective adversarial example, 
the error bars visualize the standard deviation.} \label{fig:retrainDist}
\end{figure}
Our previous analyses demonstrated the fundamental vulnerability
of classifiers to adversarial attacks.
But following the idea of "fighting fire with fire", can adversarial examples in turn also be used to protect from such attacks?
The goal of our final evaluation was to study the idea of using FGSM to protect SVM classifiers by retraining them on adversarial examples. 
We focused on the SVM classifier and the feature-level attack here because this was the most vulnerable setting.

We trained the SVM as before.
Then, we randomly chose some percentage of the training data to generate adversarial examples and trained a new SVM on the original training data and the adversarial examples.
We did not target the adversarial examples to any specific class.
As before, we report accuracy before and after the attack and calculate the distances between benign test samples and the respective adversarial samples.
Additionally, we also measured the accuracy after retraining and compare it to the accuracy without retraining to ensure the SVM was not rendered useless due to the insertion of adversarial training data.
We kept $\epsStep=0.1$ and set $\epsMax$ to 2.0, the highest perturbation we used for the previous attacks. 
Preliminary experiments showed that the percentage of training data that was added to the training dataset has the highest influence on attack success. 
We therefore evaluate for two amounts:
adding 10\% and adding 50\% of adversarial examples to the training set.

\autoref{fig:retrainAcc} summarises the results of this evaluation.
In untargeted mode and for some targeted attacks the accuracy remains between 0.4 and 0.7.
That is, in these cases the training with adversarial examples is effective in protecting the SVM classifier.
However, some targeted attacks are still successful and pull the accuracy below chance level. 
It is important to note that the accuracy after retraining remains similar to the initial accuracy (see also \autoref{tab:results}).
This shows that adding adversarial examples to the training data set does not harm the overall utility of the classifier.

When comparing distances,~\autoref{fig:retrainDist} shows that the distances between adversarial and benign data is larger due to retraining.
Especially for the targeted attack on newspaper to flip the label to comic, the attack is successful, but at a very large perturbation clearly exceeding the benign distances. 
This underlines the potential of training on adversarial examples as a first line of defence against adversarial attacks. 


\section{Discussion}
\subsection{On Adversarial Attacks}

Our evaluations highlight a serious problem that has so far been neglected in the eye tracking community:
We demonstrated that current classifiers for eye-based user modelling are not trustworthy because of their vulnerability to adversarial examples. 
For the envisioned future application of eye-based user modelling for life logging and quantified self, this means that collected gaze behaviour statistics currently can be tampered with without the user noticing it.
More seriously, for future use of eye tracking and eye-based user modelling in the medical domain~\cite{vidal2012wearable, holzman1974eye} this finding means that users currently cannot be certain that diagnostic predictions are correct before starting a respective treatment. 

Our evaluations also showed that ``security by obscurity'' is not a solution. 
Even if the type of classifier is unknown to the adversary, the classifier is still vulnerable (see \autoref{fig:fgsmAccuracy}).
We are confident that hiding the training data is not a solution either, even though we could not test this strategy in this work due to the lack of a suitable, larger dataset.
Given that eye tracking researchers increasingly publish their gaze datasets, potential attackers already have access to large amounts of training data anyway.

\subsection{On Defending against Adversarial Attacks}
Robustness of classifiers is well studied in the machine learning and security community 
and researchers have explored many attack vectors and defence mechanisms. 
It is an open question whether their findings that are predominantly on images carry over to eye data. 
As a first step to answering this question, we explored adversarial retraining. 
We found it partially successful and due to the easy generation of adversarial examples with FGSM this method is also usable in practice. 
Even better, the detailed analysis of the different document types demonstrated that some documents can be classified better in presence of adversarial perturbations. 
As a nice side effect we observed that this method does not cost utility, i.e., the classification accuracy does not drop. 
This might be simply due to the fact that more data is available, especially in those areas where the classifier would otherwise lack training data. 
Therefore, adversarial examples might also be used as a way to augment training data and generate high performance classifiers. 

\subsection{On Recommendations}
We recommend practitioners to carefully test their classifiers for vulnerabilities to adversarial examples before fully trusting their outputs. 
Especially if gradients are available for the classifiers, the overhead of generating adversarial examples from test data with FGSM is only 0.05 seconds per sample. 
Retraining with 10\% of the training data as explained in \autoref{sec:defense} takes 4-5 minutes.
Compared to under one minute for training without adversarial examples, this is a manageable overhead.
Notice that these measurements were made on a laptop without the use of multithreading.
If vulnerabilities are found, adding adversarial examples to training data might be a first line of defense that is feasible to implement.
Furthermore, we suggest eye tracking researchers to study the characteristics of adversarial examples, 
are these fundamentally different from benign data such that they can be easily detected?
Or do adversarial examples point to regions where training data is sparse and can therefore be used to augment the training set for better classification qualities?

\subsection{On Limitations}
Our evaluations provide valuable new insights into a fundamental vulnerability of current classifiers for eye-based user modelling.
Nevertheless, future work could address a number of limitations in the current study.
First, all of our findings are based on the 20-participant dataset from \cite{steil2019privacy}.
While the dataset size is not uncommon and allows for meaningful analyses, it is clear that with a view to further and potentially more sophisticated analyses, the collection of larger and more diverse datasets will be crucial.
A second limitation is in the studied reading task.
While reading is a truly pervasive activity and has therefore been studied early on in eye-based user modelling \cite{bulling08_pervasive,bulling12_tap}, it will be interesting to see whether everyday activities are more diverse and corresponding classifiers will therefore be harder to attack. 
Third, our attack on raw data was tested only with one common eye movement event detection algorithm.
Future work could explore whether other event detection methods, e.g., velocity-based instead of dispersion, can be more robust to adversarial attacks. 
Finally, future work could study the applicability of other defense mechanisms to eye tracking classifiers than the adversarial training proposed here, such as defensive distillation~\cite{papernot2016distillation}.


\section{Conclusion}
In this work we studied adversarial examples -- a novel attack vector in the emerging field of privacy-aware eye tracking.
Specifically, we demonstrated the vulnerability of two commonly used classifiers for eye-based user modelling to adversarial examples on the sample task of eye-based document type classification.
The identified vulnerabilities underline the urgent need for further research on more robust gaze data classifiers for eye-based user modelling.
As a first step in this direction, we demonstrated that training with a small amount of adversarial examples can be effective for hardening SVM classifiers against adversarial attacks.
\bibliographystyle{ACM-Reference-Format}
\bibliography{eyetracking.bib}

\appendix
\begin{table*}[t]
\begin{tabular}{llc|rrlrlrl}
\toprule
metric &attack type & target & untargeted & \multicolumn{2}{c}{comic classified as} & \multicolumn{2}{c}{newspaper classified as} & \multicolumn{2}{c}{textbook classified as} \\ 
\midrule
& &  & & newspaper & textbook & comic & textbook & comic & newspaper \\ 
\hline
feature level attacks \\
accuracy &  initial & SVM & 0.89& 0.90& 0.90& 0.87& 0.87& 0.87& 0.87\\ 
 & general $\epsMax$ & SVM & 0.24& 0.08& 0.03& 0.02& 0.40& 0.02& 0.00\\ 
 & individual $\epsMax$ & SVM & 0.24& 0.07& 0.03& 0.02& 0.40& 0.02& 0.00\\
\hline
accuracy & initial & RF & 0.86& 0.91& 0.91& 0.84& 0.84& 0.82& 0.82\\ 
& general $\epsMax$ & RF & 0.43& 0.09& 0.45& 0.23& 0.58& 0.51& 0.38\\ 
& individual $\epsMax$ & RF & 0.43& 0.09& 0.44& 0.23& 0.58& 0.48& 0.38\\  
\hline
 distance  &   benign & SVM & 7.76& 5.94& 5.94& 5.42& 5.42& 4.59& 4.59\\ 
 &   individual $\epsMax$& SVM & 4.32& 3.18& 5.34& 7.99& 2.55& 4.93& 1.75\\ 
 &   general $\epsMax$& SVM & 4.35& 3.27& 5.34& 8.89& 2.58& 4.97& 1.75\\ 
 &   guess & SVM & 3.73& 2.49& 5.20& 7.51& - & 3.45& 1.67\\ 
\hline
  distance  &   benign & RF & 7.76& 5.94& 5.94& 5.42& 5.42& 4.59& 4.59\\ 
 &   individual $\epsMax$& RF & 3.97& 2.95& 5.24& 8.42& 2.45& 4.25& 1.75\\ 
 &   general $\epsMax$ & RF & 4.35& 3.08& 5.34& 8.89& 2.58& 4.97& 1.75\\ 
 &   guess & RF & - & 2.49& - & 7.51& - & -& -\\ 
\hline
retrain attacks \\
accuracy &   10\% initial & SVM &0.89& 0.95& 0.95& 0.86& 0.86& 0.84& 0.84\\ 
&   10\% retrain  & SVM &0.88& 0.88& 0.88& 0.88& 0.88& 0.88& 0.88\\ 
&   10\% attack & SVM &0.46& 0.30& 0.01& 0.00& 0.41& 0.66& 0.05\\ 
accuracy &   50\% initial & SVM &0.89& 0.94& 0.94& 0.84& 0.84& 0.81& 0.81\\ 
&   50\% retrain  & SVM &0.86& 0.86& 0.86& 0.86& 0.86& 0.86& 0.86\\ 
&   50\% attack & SVM &0.56& 0.53& 0.16& 0.00& 0.49& 0.68& 0.07\\ 
\hline 
 distance &   10\% benign & SVM &7.76& 5.94& 5.94& 5.42& 5.42& 4.59& 4.59\\ 
&   10\% attack & SVM &7.20& 6.48& 4.54& 15.56& 2.91& 4.23& 2.44 \\
 distance &   50\% benign & SVM &7.76& 5.94& 5.94& 5.42& 5.42& 4.59& 4.59\\ 
&   50\% attack & SVM &7.47& 7.10& 3.94& 16.67& 2.75& 4.54& 2.39 \\
\bottomrule
\end{tabular} 
\caption{Summaries of results for the different types of attacks
} \label{tab:results}
\end{table*}

\begin{table*}[t]
\begin{tabular}{l|l|r}
\toprule
feature type & explanation & number of features \\
\midrule
Fixation & rate, computed over pupil positions within one fixation: mean, max, variance of durations, \\
& mean of mean, variance of variance & 8 \\
Saccades & rate, ratio of (small/large/right/left) saccades, mean, max and variance of amplitudes & 12\\
Combined & ratio saccades to fixations & 1\\
\hline
Wordbooks & for n-grams of length up to four (including): \\
	& number of non-zero entries, maximum and minimum of entries & 24\\
\hline
Blinks & rate, mean and variance of blink duration & 3\\
Pupil Diameter & mean of mean and variance of variance during fixations & 4\\
\midrule
reading features & euclidean distance between the 5\% and 95\% quantile of fixation coordinates, \\
& the slope of saccade directions using linear regression over fixations & 2 \\
\bottomrule
\end{tabular}
\caption{We extracted 54 eye movement features to describe a user's eye movement behaviour based on the 52 features by Steil et al.~\cite{steil2019privacy} and 2 features from Kunze et al.~\cite{kunze2013know}}    \label{tab:features}
\end{table*}
\end{document}